\documentclass[Afour,sageh,times]{sagej}
\usepackage{moreverb,url}
\usepackage[colorlinks,bookmarksopen,bookmarksnumbered,citecolor=red,urlcolor=red]{hyperref}
\usepackage{graphicx}
\usepackage{multirow}
\usepackage{caption}
\usepackage{subcaption}
\usepackage{adjustbox}
\usepackage{natbib}
\usepackage{pdflscape}
\newcommand\BibTeX{{\rmfamily B\kern-.05em \textsc{i\kern-.025em b}\kern-.08em T\kern-.1667em\lower.7ex\hbox{E}\kern-.125emX}}
\def\volumeyear{2024}

\begin{document}
\runninghead{Dhabu \textsc{et al}}
	
\title{Characterizing Rotational Ground Motions: Implications for Earthquake-Resistant Design of Bridge Structures}
	
\author{Anjali C. Dhabu\affilnum{1} Felix Bernauer\affilnum{2} Chun-Man Liao\affilnum{3} Ernst Niederleithinger\affilnum{3} Heiner Igel\affilnum{2} Celine Hadziioannou\affilnum{1}}
	
\affiliation{\affilnum{1}Institute of Geophysics, Centre for Earth System Research and Sustainability (CEN), Universit\"{a}t Hamburg, Hamburg, Germany\\
\affilnum{2}Department of Earth and Environmental Sciences, Ludwig-Maximilians University, Munich, Germany \\
\affilnum{3} Division of Non-Destructive Testing Methods for Civil Engineering, Bundesanstalt f\"{u}r Materialforschung und-pr\"{u}fung, Berlin,Germany}
	
\corrauth{Anjali Dhabu, Institute of Geophysics, Centre for Earth System Research and Sustainability (CEN), Universit\"{a}t Hamburg, Hamburg, Germany}
	
\email{anjali.dhabu@uni-hamburg.de}
	
\begin{abstract}
	Earthquakes cause catastrophic damage to buildings and loss of human life. Civil engineers across the globe design earthquake-resistant buildings to minimize this damage. Conventionally, the structures are designed to resist the translational motions caused by an earthquake. However, with the increasing evidence of rotational ground motions in addition to the translational ground motions due to earthquakes, there is a crucial need to identify if these additional components have an impact on the existing structural design strategies. In this regard, the present study makes a novel attempt to obtain the dynamic properties of a large-scale prototype prestressed reinforced concrete bridge structure using six component (6C) ground motions. The structure is instrumented with conventional translational seismic sensors, rotational sensors and newly developed six-component sensors under operating and externally excited conditions. The recorded data is used to carry out Operational Modal Analysis and Experimental Modal Analysis of the bridge. Modal analysis using the rotational measurements shows that the expected location of maximum rotations on the bridge differs from the maximum translations. Therefore, further understanding the behavior of rotational motions is necessary for developing earthquake-resistant structural design strategies.
\end{abstract}
	
	
\maketitle

\section{Introduction}\label{sec1}
Earthquakes cause catastrophic damage to infrastructure and loss of human life, leading to heavy financial derailment of an economy. To curtail these damages, designing earthquake-resistant infrastructure has been the main aim of structural engineers for the last few decades (\cite{Fajfar1997, Chandler2001, Agarwal2006}).  Modal analysis is the most used tool to obtain information regarding the natural frequencies and mode shapes of the structure (\cite{Kunnath2004, Brincker2014}). When excited at these natural frequencies, the structures are expected to undergo maximum vibration, and the mode shapes indicate the regions of maximum displacement and, in turn, maximum damage at these natural frequencies. As normally only translational ground motions due to earthquakes are measured due to earthquakes, it is customary to carry out operational modal analysis (OMA) or experimental modal analysis (EMA) using translational data (\cite{Gentile2014}) to model and understand the dynamic behavior of a structure. 

In seismology, it is now well known that earthquakes also induce rotational ground motions (\cite{Igel2005}). This development raises important questions to civil engineers: (i) Are the structures subjected to additional forces due to rotational ground motions? (ii) Does the collapse mechanism of the structure change in the presence of additional forces? (iii) Are existing design strategies adequate to withstand these movements? This paper aims to address these questions by analyzing the dynamic behavior of structures using the six-component translational and rotational records. Estimating eigenfrequencies and eigenmodes derived from both translational and rotational motions can offer valuable insights into the significance of rotational motions and their integration into earthquake-resistant design strategies. Post-earthquake assessments have frequently revealed damage patterns resulting from torsional forces, even in buildings with symmetric layouts (\cite{Celebi2006, Michel2009}). Such torsional forces are perceived to arise from asymmetries in the building design or accidental asymmetries introduced by discrepancies between theoretical and actual designs (\cite{Newmark1969, Kan1977, Kan1981, Chandler1986, Fajfar1992}). However, it remains intriguing to investigate whether these torsional forces stem from the building design itself or are inherent to the seismic waves. Understanding the nature of the seismic forces to which buildings are subjected during earthquakes is crucial for developing effective earthquake-resistant design strategies. 

Prior to the development of sophisticated sensors, the rotational ground motions were estimated mathematically as the curl of the recorded translational motions. \cite{Bouchon1982} found that the magnitude of the rotational ground motions was too low to damage any structure during an earthquake. However, as the rotation sensors were not available then, it was difficult to validate this hypothesis. \cite{nigbor1994} was the first to develop a rotation measuring sensor and used it to record the 6C ground motions due to an explosive event. Using the same instrument, \cite{Takeo1998} recorded the 6C ground motion data for the M$_w$ $5.2$ magnitude Izu peninsula earthquake on November 26, 1998, in Japan. They compared the recorded rotational ground motions with the analytically obtained rotations for the Earth medium and found that the analytical solutions drastically underestimated the actual rotation amplitude. Eventually, Ring laser gyroscopes, primarily developed to record variations in the Earth's absolute rotation rate, successfully measured rotational ground motions due to M$_w$ $8.1$ magnitude Tokachi-Oki earthquake, M$_w$ $7.0$ New Ireland earthquake, M$_w$ $7.3$ magnitude Vanuatu earthquake etc. (\cite{Igel2005, stedman1995, pancha2000, Suryanto2006, cochard2006}). The Eentec R-1 rotational seismometer was the first portable and modestly priced sensor to record strong motion from small earthquakes of magnitude up to approximately M$_w$ $4$ at distances up to several tens of kilometers (\cite{Lee2009, Lee2009a, nigbor2009, Wassermann2009, Bernauer2012}). Thereafter, iXBlue, a France-based corporation, in collaboration with the European Research Council Project, ROMY (Rotational Motions - A new observable for Seismology), for the first time developed a three-component interferometric fiber optic gyroscope (IFOG), named, BlueSeis-3A. These sensors are suitable for field applications in seismology, volcanology, ocean bottom seismometers, and earthquake engineering (\cite{Bernauer2016, Bernauer2018}) . In the recent past, \cite{MurrayBergquist2021} carried out a laboratory investigation on a set of $20$ IMU50 sensors developed by iXblue, to find its applicability in the field of seismology and civil engineering. IMU50 is a collocated 6-degree-of-freedom (6-DOF) sensor that measures translational and rotational motions. These sensors use three perpendicular silicon-based capacitive microelectromechanical system (MEMS) accelerometers to measure translational motions and three fibreoptic gyroscopes (FOG) coiled around the translational axes to measure the rotational motions. This arrangement makes the sensor very compact, small in size, easy to mount and install on civil engineering structures, and appropriate for structural health monitoring (SHM) purposes. In the experimental investigations carried out by \cite{MurrayBergquist2021}, it was concluded that the IMU50 sensors are not suitable for recording ambient seismic noise because of the relatively higher self-noise of the sensor. However, they can be instrumental in recording the response of buildings during an external excitation. 

As the development of rotational sensors gained importance, many researchers analyzed the importance of rotational data in various domains of seismology (\cite{Wassermann2020, Keil2020, Kurzych2020, Izgi2021, Wassermann2022}) and civil engineering. \cite{Lin2012} were the first to compare array derived rotations (ADR) from translational seismometers with rotation measured by an Eentec R1 sensor in the 101 Taipei tower, Taiwan. They concluded that the accuracy of the ADR about the orthogonal components in the horizontal plane is affected by the array configuration. \cite{Gueguen2020} analyzed the rotation rates derived from a network of translation sensors and those obtained from BlueSeis-3A installed at the Grenoble city hall building. They compared the ADR to those provided by the rotation sensor and showed that the calculation of rotation in structures using station pairs aligned in the main directions underestimates the rotation rates. They also highlight that the interpretation and analysis of rotations recorded along the three axes in structures under earthquake excitation are essential to developing earthquake engineering design codes. Recently \cite{Dzubay2022} instrumented the top of a slender rock tower located near Moab, Utah, USA, and recorded ambient vibration data from translational and rotational seismometers. They concluded that the data obtained from translational sensors provided information on the eigenfrequency and eigenvector, while the rotational sensors revealed information about the eigenmodes and orientations of the modal rotation axes. \cite{Rossi2023} demonstrated that an accelerometer and rotational sensor collocated at a single station can be used to describe the dynamic properties of a high-rise building and obtained the first 20 eigenfrequencies. They concluded that a single set of 6-DOF sensors installed on its roof is sufficient to identify tall buildings' modal frequencies and mode shapes. 

It is essential to note that the response of structures like tall buildings and free-standing rock towers is expected to remain similar when exposed to translational or rotational motions due to inherent boundary conditions at the supports.  However, structures supported at more than one location, like culverts, bridges, stadium roofs, etc., are expected to show different responses when subjected to translational and rotational motions. As a first step to understand this difference, the present paper makes an attempt to determine the eigenfrequencies and eigenmodes of a civil engineering structure under ambient noise conditions and external excitation using both translational and rotational motions. For this purpose, an experiment is carried out on BLEIB, a $24$m long large-scale prototype bridge model built by the Bundesanstalt f{\"u}r Materialforschung und pr{\"u}fung (BAM), Berlin, Germany, for experimental purposes. In the present work, this prototype bridge is instrumented with four conventional translational seismometers (Trillium Compact) and four rotational seismometers (BlueSeis-3A) to measure translations and rotations on the bridge for 18 days. In addition, four 6-DOF sensors (IMU50) were also installed on the bridge for two days to measure translational and rotational motions due to the non-destructive active experiments carried out on the bridge. The active experiments included (i) changing the prestress of the bridge from $450$kN to $200$kN and back to $450$kN (ii) placing additional load on the bridge up to $900$kg and (iii) hammer hits.  The primary aim of the experiment is to understand the response of the bridge under (i) ambient noise conditions and (ii) active conditions and assess the usability of IMU50 sensors for structural health monitoring purposes. The recorded translational and rotational data from all three types of sensors is analyzed to determine the modal frequencies and mode shapes of the structure.   
 
\section{The Experimental Setup}\label{sec2}
The present study uses a large-scale prototype of the prestressed concrete simply-supported continuous bridge, BLEIB,  available at the Bundesanstalt f{\"u}r Materialforschung und pr{\"u}fung (BAM), Berlin, Germany, to carry out the experiments. Figure \ref{fig:photos of bridge}(a) and \ref{fig:photos of bridge}(b) shows the prototype bridge, which is $24$m long and $0.9$m wide, constructed using M45 grade concrete with standard steel reinforcements.  
\begin{figure*}[htb!]
	\begin{center}
		\includegraphics[width = 0.9\textwidth]{Figure1.pdf}
	\end{center}
	\caption{\textbf{(a,b)}: Experimental Setup of the BLEIB structure. \textbf{(c)}: Schematic sketch of the BLEIB structure showing the location of instruments, loads, and hammer hits on the bridge. At Station 1, Station 2, Station 3 and Station 4, the structure is instrumented with 3-DOF TC (Trillium Compact) and BlueSeis3A (BS) sensors that measure translations and rotations respectively. The 6-DOF IMU50 sensors are placed at S1, S2 and S3 while S4 location is instrumented with 6-DOF LCG demonstrator.}
	\label{fig:photos of bridge}
\end{figure*}
It has two spans of $12$m each. The bridge is further fixed with steel bars connected to hydraulic jacks on both ends to provide prestress in the bridge. The bridge is instrumented in two phases (i)passive and (ii) active phase during the experiment. During the passive experiments, the bridge is instrumented with Trillium compact (TC) and Blueseis3A (BS) sensors at four locations marked as Station 1, 2, 3 and 4 in Figure \ref{fig:photos of bridge}(c). Trillium compact (TC) (Figure \ref{fig:instruments}(a)) is a broadband seismometer that measures the translational velocity along the X- Y- and Z- axes, and BlueSeis (BS)-3A (Figure \ref{fig:instruments} (b)) measures the rotation rate about the X-, Y-, and Z-axes.  These stations are located at a distance of $6$m from each other. Station 1 is located at a distance of $3.5$m from the right support, while Station 4 is located at a distance of $2.5$ m from the left support. Stations 2 and 3 are located at $2.5$m and $3.5$m, respectively, from the intermediate support. During the active experiments, in additon to TC and BS sensors, the bridge is instrumented with IMU50 sensors (Figure \ref{fig:instruments}(c)) and LCG demonstrator (Figure \ref{fig:instruments}(d)) which measure both translational acceleration and rotation rate. IMU50 sensors are placed at three locations shown as `S1', `S2', and `S3' in Figure \ref{fig:photos of bridge}(c) and LCG demonstrator is place at location `S4' (Figure \ref{fig:photos of bridge}(c)). S1 and S3 are located near to the right and intermediate support at a distance of $1$m and $0.6$m respectively while S2 and S4 are located at a distance of $5.4$m and $5$m from the left and intermediate support respectively. All instruments' sampling frequencies except the LCG demonstrator are $200$Hz. The sampling frequency of the LCG demonstrator is $100$Hz. It should be noted here that TC and BS are placed on the bridge for the whole duration of the experiment, while the IMU50 sensors and LCG demonstrator are placed only on the days when active experiments are carried out on the bridge.

\begin{figure*}[htb!]
    \centering
    \includegraphics[width = 0.9\textwidth]{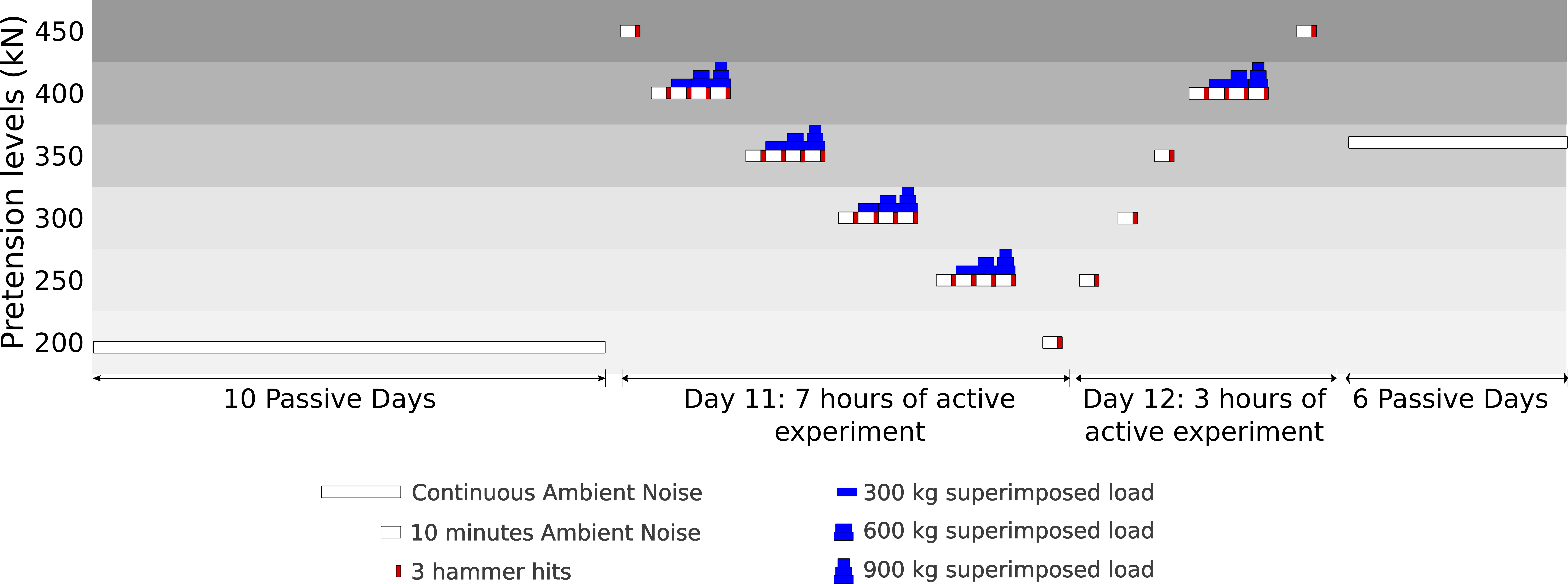}
    \caption{Schematic sketch to show the experiments carried out on the BLEIB structure over a period of 18 days}
    \label{fig:experiment_sketch}
\end{figure*}

Figure \ref{fig:experiment_sketch} schematically describes the experimental regime that was followed for the BLEIB structure.
The bridge is instrumented for a total duration of 18 days, starting from September 27, 2021, 07:50:00 UTC to October 15, 2021, 07:01:51 UTC, out of which the bridge was in a passive state for 16 days, i.e., it was monitored under operating conditions. From September 27, 2021, to October 6, 2021, a prestress of $200$kN was maintained on the bridge, and the data was recorded continuously by BS and TC sensors. Thereafter, for two days, October 7 and October 8, 2021, active experiments were carried out on the bridge. The experiments included (i) change in the prestress levels of the bridge, (ii) applying additional load, and (iii) impactful hammer hits. During the active experiments, in addition to the BS and TC sensors, 3 IMU50 sensors and 1 LCG demonstrator were also installed on the bridge (Figure \ref{fig:photos of bridge}(c)). The active experimental regime started by increasing the prestress of the bridge from $200$kN to $450$kN over a duration of $40$ minutes. At this prestress level, $10$ minute of ambient seismic data was recorded followed by three consecutive hammer hits as shown in Figure \ref{fig:experiment_sketch}. The hammer hits were provided between Station 3 and Station 4 at a distance of $4.7$m from the left support (Figure \ref{fig:photos of bridge}(c)). Thereafter, the prestress level of the bridge was decreased to $400$kN, at which data was recorded for $10$ minutes of ambient noise and three consecutive hammer hits. At the same prestress level, an additional load of $300$kg was applied on the bridge between Station 3 and Station 4 at a distance of $6.5$m from the left support. After adding the superimposed load, ambient data is recorded for $10$ minutes, and three consecutive hammer hits are applied. Thereafter, the process is repeated for superimposed loads of $600$kg and $900$kg. In a similar manner, the prestress is reduced in steps of $50$kN each up to $200$kN and then increased back to $450$kN. Every time the prestress was changed, or a superimposed load was added to the bridge, an initial $10$ minutes of ambient noise was recorded, followed by hammer hits. After the active experiments, the bridge was in the passive state again from October 8, 2021, 14:30:00 hours to October 15, 2021, 07:01:51 hours and a prestress of $360$kN was maintained on the bridge. The TC and BS sensors recorded the ambient data during these days. Finally, at the end of the experiment, a total of $24\times16$ hours of passive data and two days of active data was available for analysis. 
   
\section{Analysis of Experimental Data}
The first step towards designing a structure is to identify its natural frequencies and determine its deflected shape when excited at these natural frequencies. For a structure supported and fixed at two ends that can be represented by $n$ masses, $m_i$ lumped at a regular interval and connected by $n+1$ springs of stiffness coefficient, $k_i$, and dampers of damping coefficient $c_i$ with no external forces applied on the system, the equation of motion can be written as:
\begin{equation}
    \left[M\right]\Ddot{u}(t)+\left[C\right]\dot{u}(t) +\left[K\right]u(t)=0
    \label{eq:EOM}
\end{equation}
where $\left[M\right] = \begin{bmatrix}
        m_1 & 0 & ... & ... & 0\\
        0 & m_2 & ... & ... & 0\\
        0 & : & : & : & 0\\
        0 & : & : & m_{n-1} & 0\\
        0 & : & : & : & m_n
    \end{bmatrix} $, $\left[C\right] = \begin{bmatrix}
        c_1 & (c_1+c_2) & ... & ... & 0\\
        c_2 & (c_2+c_3) & c_3 & ... & 0\\
        0 & c_3 & : & : & 0\\
        0 & : & : & (c_{n-1}+c_n) & c_n\\
        0 & : & : & c_n & (c_n+c_{n+1})
    \end{bmatrix}$,\\ $\left[K\right] = \begin{bmatrix}
        k_1 & (k_1+k_2) & ... & ... & 0\\
        k_2 & (k_2+k_3) & k_3 & ... & 0\\
        0 & k_3 & : & : & 0\\
        0 & : & : & (k_{n-1}+k_n) & k_n\\
        0 & : & : & k_n & (k_n+k_{n+1})
    \end{bmatrix}$ and 
    $\left\{u\right\} = \begin{Bmatrix}
        u_1 \\
        u_2\\
        :\\
        u_{n-1}\\
        u_n
    \end{Bmatrix}$\\
\\
In the above equation, $\Ddot{u}(t)$, $\dot{u}(t)$, and $u(t)$ represent the acceleration, velocity, and displacement vector, respectively. It is assumed in equation\eqref{eq:EOM} that the masses in the system have a degree of freedom only in one direction and can be solved analytically by assuming appropriate solutions for the displacement to determine the eigenfrequencies and eigenmodes. However, in the real structures, it is difficult to employ analytical solutions for two reasons: (i) it is difficult to practically represent a real structure in the form of lumped masses, and (ii) the masses are subjected to three/six degrees of freedom rather than just one. Therefore, to avoid this complexity, numerical techniques are explored in order to carry out the modal analysis. The modal parameters (natural frequencies, mode shapes, damping, etc.) of a structure can be estimated using many methods like (i) Peak Picking (PP), (ii) Frequency Domain Decomposition (FDD), (iii) Enhanced Frequency Domain Decomposition (EFDD), (iv) Frequency Spatial Domain Decomposition (FSDD), (v) Covariance driven Stochastic Subspace Identification (CDSSI), and (vi) Data-driven Stochastic Subspace Identification (DDSSI). In the FDD method, if there are `$r$' number of unknown inputs, $x(t)$ and `$m$' number of measured responses, $y(t)$, then the Power Spectral Density (PSD) matrix of the inputs, $G_{xx}(\iota\omega)$ is related to the PSD matrix of the responses, $G_{yy}(\iota\omega)$ as (\cite{Hizal2020}): 
\begin{equation}
    G_{yy}\left(\iota\omega\right) = \overline{H}(\iota\omega)G_{xx}(\iota\omega)H(\iota\omega)^T
\end{equation}
where $H(\iota\omega)$ is the $m\times r$ Frequency Response Function (FRF) matrix, and $\overline{H}(\iota\omega)$ and $H(\iota\omega)^T$ represents its complex conjugate and transpose respectively. In the residual form, the Frequency Response Function (FRF) can be written as:
\begin{equation}
    H(\iota\omega) = \sum_{k=1}^n \frac{R_k}{\iota\omega -\lambda_k} + \frac{\overline{R}_k}{\iota\omega -\overline{\lambda}_k}
\end{equation}
where $n$ is the number of nodes, $\lambda_k$ is the pole and $R_k$ is the residue given by 
\begin{equation}
    R_k = \phi_k\gamma_k^T
\end{equation}
where $\phi_k$ is the mode shape vector and $\gamma_k$ is the modal participation vector. The EFDD and FSDD method are further modifications of the FDD method for accurate determination of the closely spaced eigenfrequencies and damping ratios. 

\subsection{Operational Modal Analysis}
Operational Modal Analysis (OMA), also known as ambient modal identification aims to determine the modal properties of the structure using the vibration data recorded under the structure's operating conditions. In this analysis, it is assumed that the structure is not subjected to any initial excitation or any known artificial excitation. As discussed in the previous section, the bridge was instrumented and monitored for a total duration of 18 days and translational and rotational data was recorded continuously during this period. Amongst the 18 days, the bridge was under operating conditions for 16 days on which the BS and TC sensors recorded the ambient data. The recorded data for every $24$ hours was divided into the day and night time where day time is considered from 07:00:00 UTC to 17:59:59 UTC while the night is considered from 18:00:00 UTC to 06:59:59 UTC on the following day. The data is detrended and bandpass filtered in a frequency range of $0.5$ Hz to $40$ Hz. 
\subsubsection{Eigenfrequencies}
\begin{figure*}[htb!]
\centering
    \begin{subfigure}[b]{0.45\textwidth}
         \includegraphics[width=\textwidth]{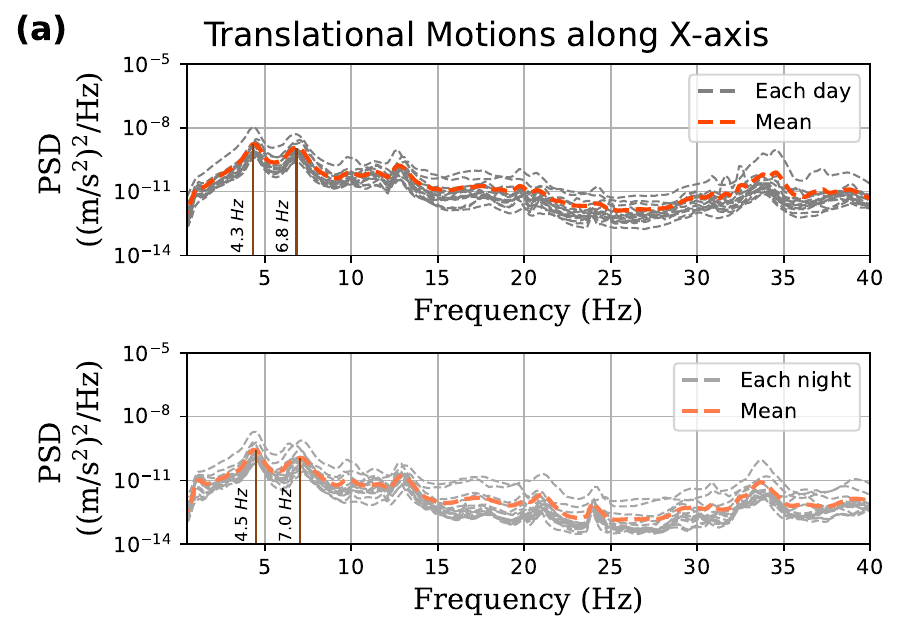}
         \label{BBAM4-X}
     \end{subfigure}
     \begin{subfigure}[b]{0.45\textwidth}
         \includegraphics[width=\textwidth]{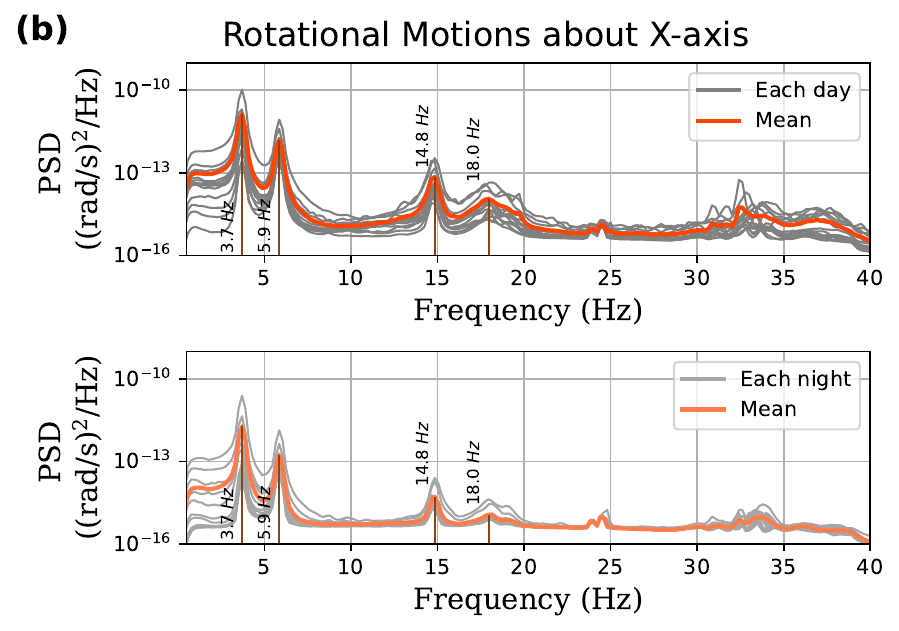}
         \label{RBAM4-X}
     \end{subfigure}\\
     \begin{subfigure}[b]{0.45\textwidth}
         \centering
         \includegraphics[width=\textwidth]{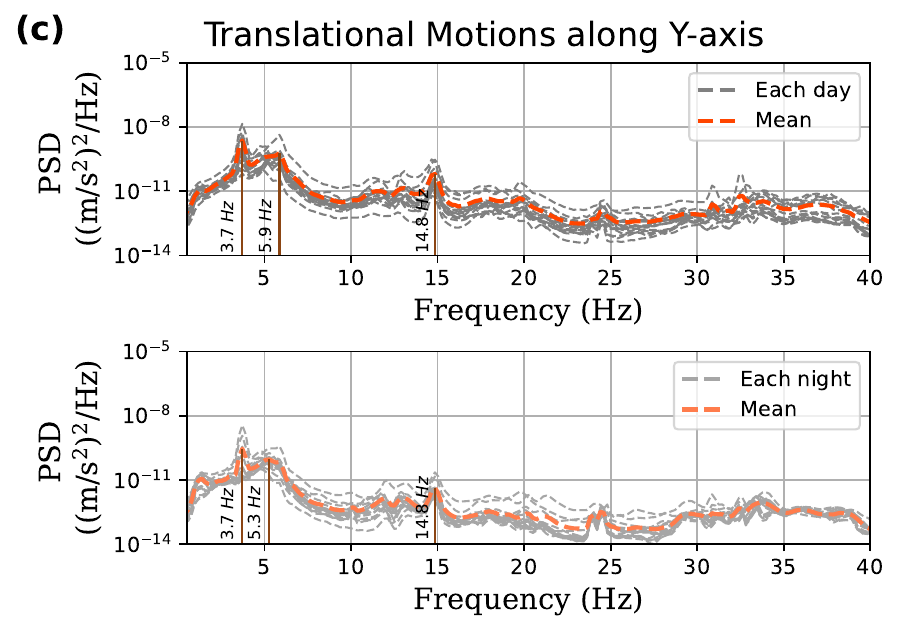}
         \label{BBAM4-Y}
     \end{subfigure}
	 \begin{subfigure}[b]{0.45\textwidth}
	 	\centering
	 	\includegraphics[width=\textwidth]{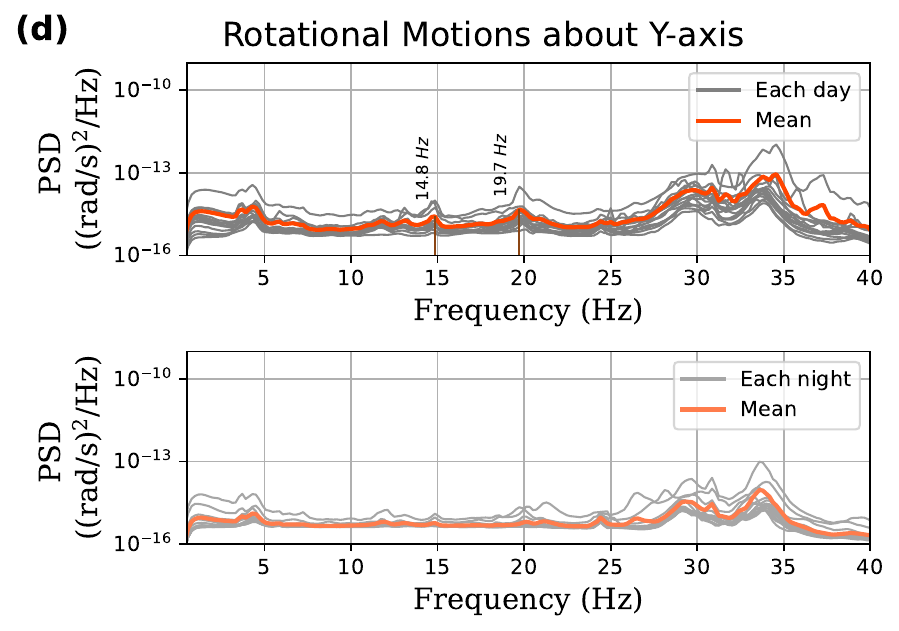}
	 	\label{RBAM4-Y}
	 \end{subfigure}\\
	 \begin{subfigure}[b]{0.45\textwidth}
	 	\centering
	 	\includegraphics[width=\textwidth]{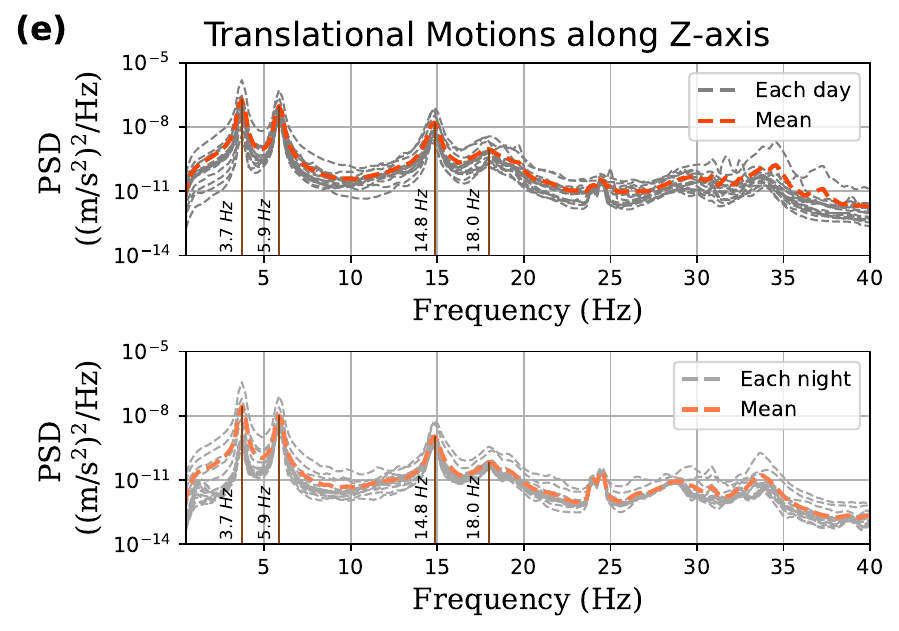}
	 	\label{BBAM4-Z}
	 \end{subfigure}
	 \begin{subfigure}[b]{0.45\textwidth}
	 	\centering
	 	\includegraphics[width=\textwidth]{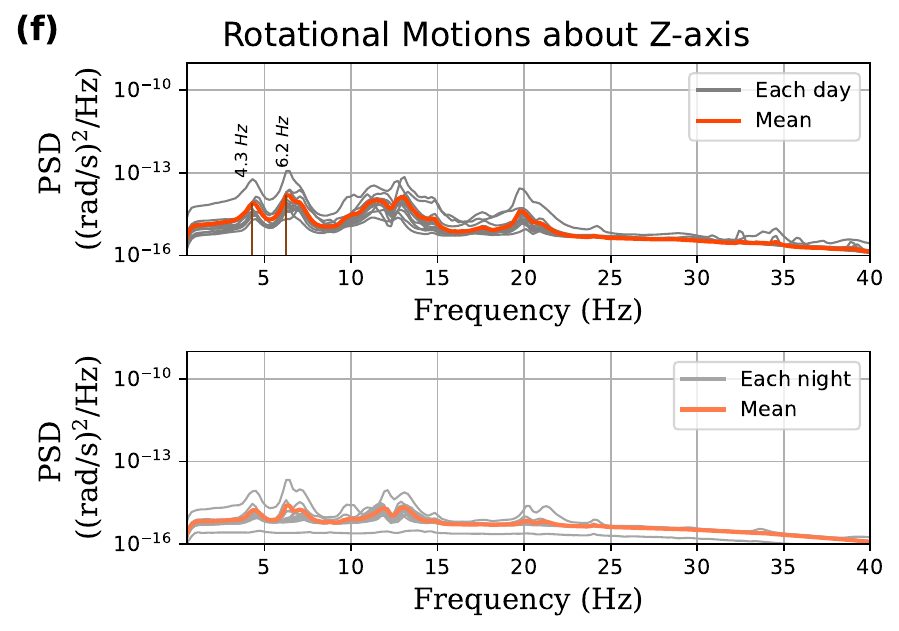}
	 	\label{RBAM4-Z}
	 \end{subfigure} 
     \caption{Variation of Power spectral density (PSD) w.r.t the frequency for ambient data recorded in the passive days (15 days and 16 nights) of the experiment. \textbf{(a), (c), and (e)}: Variation of PSD of translational acceleration recorded by the Trillium Compact sensors in the \textit{x-}, \textit{y-} and \textit{z-} directions respectively w.r.t. Frequency at Station 4. \textbf{(b), (d), and (f)}: Variation of PSD of rotational rate recorded by BS sensors about the \textit{x-}, \textit{y-} and \textit{z-} directions respectively w.r.t Frequency at Station 4 \\(Note: In each subplot the top-panel shows the day data and the bottom panel shows the night data. The grey lines show the variation of PSD wrt frequency for each day/night and the red line signifies the average PSD variation w.r.t frequency).}
    \label{fig:PSD_all}
\end{figure*}

To identify the natural frequencies, the power spectral density (PSD) of the recorded data was calculated at all frequencies. For data recorded in the time domain $X\left(t\right)$, its power spectral density function $X_{PSD}\left(f\right)$ is given as 
\begin{equation}
    X_{PSD}\left(f\right) =\lim_{\Delta f \to 0} \left[\frac{1}{2}\frac{X\left(f\right)X^{*}\left(f\right)}{\Delta f}\right]
\end{equation}
where $X\left(f\right)$ is discrete Fourier transform of $X\left(t\right)$. Figure \ref{fig:PSD_all}(a)-\ref{fig:PSD_all}(f) represent the variation of PSD with respect to (w.r.t) frequency for the translational and rotational data recorded by the TC and BS sensors, respectively, at Station 4 for 15 passive days and 16 nights of the experiment. The data recorded on the first day of the experiment is neglected in the analysis due to human activity around the experiment site.  
It can be observed from Figure\ref{fig:PSD_all}(a) and \ref{fig:PSD_all}(c) that, as expected, the PSD amplitude of translational data recorded by the TC sensor is very negligible along the \textit{x-} and \textit{y-} axes respectively. It should be noted here that the \textit{x-} axis is along the lateral direction of the bridge while the \textit{y-} axis is along the longitudinal direction (Figure \ref{fig:photos of bridge}(c)). However, along the \textit{z-} axis, which is the vertical axis of the bridge, the amplitude of PSD (Figure \ref{fig:PSD_all}(e)) is approximately $1000$ times that of the PSD in the other two directions. This observation follows the expected behavior of the bridge, as it shows dominant bending movement in the vertical direction under steady-state conditions. It is also observed from Figure\ref{fig:PSD_all}(e) that under ambient noise conditions, the maximum energy of the system is concentrated at a frequency of $3.7$Hz, $5.9$Hz, and $14.8$Hz. These frequencies are identified as the first three modal frequencies of the bridge structure. Though not distinct, the fourth peak can be observed at a frequency of $18$Hz. Figure \ref{fig:PSD_all}(a) and \ref{fig:PSD_all}(c) indicate that the recorded data in the non-dominant direction is not able to capture all the natural frequencies of the bridge. It is also observed from the figures that the natural frequencies in the \textit{x-} direction are slightly higher as compared to those obtained in the \textit{y-} and \textit{z-} direction due to the higher stiffness of the structure along the \textit{x-} axis. A slight difference observed in the peak frequencies obtained from the day and night data along the \textit{x-} direction and \textit{y-} direction is attributed to the temperature and wind variation during the day and night time.  

While analyzing the rotational data recorded by the BS sensor, high amplitudes of PSD are observed for the rotation rate recorded about the \textit{x-} axis as compared to those observed about the \textit{y-} and \textit{z-} axes (Figure \ref{fig:PSD_all}(b), \ref{fig:PSD_all}(d) and \ref{fig:PSD_all}(f)). The rotation about the lateral direction (\textit{x}-axis) of the bridge shall result in the displacement in the vertical direction. As it is the dominant direction of motion, it is expected to observe larger rotations about the lateral direction as compared to the other two axes. $3.5$Hz, $5.9$Hz, $14.8$Hz, and $18$Hz are identified as the first four modal frequencies of the system, which are consistent with those observed from the translational sensors (Figure \ref{fig:PSD_all}(e)). It can also be observed from Figure \ref{fig:PSD_all}(b), \ref{fig:PSD_all}(d) and \ref{fig:PSD_all}(f) that the noise levels reduce drastically during the night time suggesting that the conventional rotational sensors are susceptible to noise. Also, more significant levels of disturbances are observed in the variation of PSD w.r.t. frequency for rotational motions about the $y-$ axis (Figure \ref{fig:PSD_all}(d)) at frequencies between $30$Hz to $35$Hz which are indicative of the higher modes of the system. It is also observed that as the stiffness of the bridge is higher in the lateral direction, slightly higher modal frequencies are captured from the rotational data about the \textit{z-}axes which is consistent with the observation made from translational data recorded along the \textit{x-}axes. 

\begin{figure}[htb!]
	\centering
    \begin{subfigure}{0.5\textwidth}
         \includegraphics[width=\textwidth]{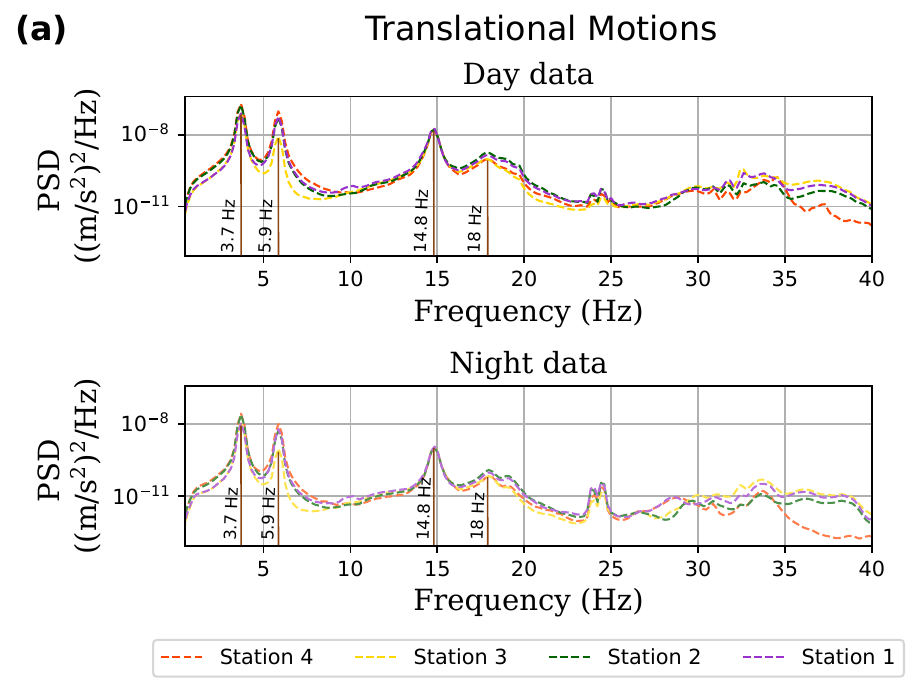}
     \end{subfigure}\\
     \begin{subfigure}{0.5\textwidth}
         \includegraphics[width=\textwidth]{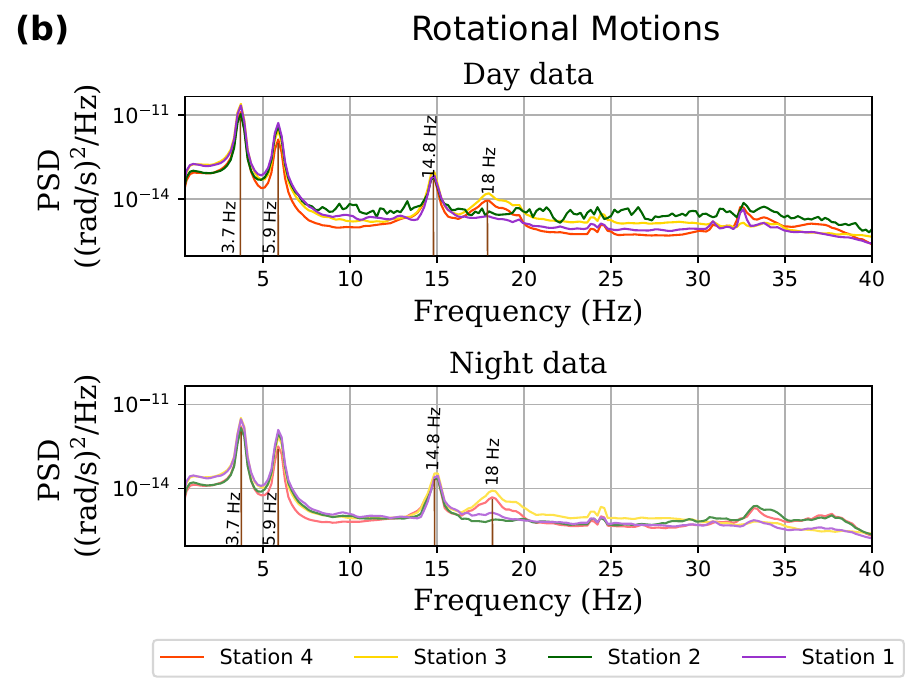}
     \end{subfigure}
    \caption{Variation of mean PSD observed at the four stations w.r.t the frequency. \textbf{(a)}: translational data recorded by the TC sensor along the \textit{z-} axis (top panel: day data and bottom panel: night data) and \textbf{(b)}: rotational data recorded by the BS sensor about the \textit{x-} axis (top panel: day data and bottom panel: night data).}
    \label{fig:PSD_all_stations}%
\end{figure}

It is evident from Figure \ref{fig:PSD_all} that the translational data recorded along the \textit{z-} axis and rotational data recorded about the \textit{x-} axis shall provide maximum insight into the behavior of the bridge. Therefore, this data is further analyzed for all four stations on the bridge. Figure \ref{fig:PSD_all_stations}(a) and \ref{fig:PSD_all_stations}(b) represents the variation of mean PSD w.r.t the frequency for the translational and rotational data respectively recorded at the four stations during the passive days of the experiment.  
It can be observed from the figure that consistent peak PSD amplitudes (modal frequencies) are obtained for the four stations from the translational and rotational data. The TC sensor evidently captures the fourth modal frequency at $18$Hz. However, in the case of rotational data, it is captured well only by BS sensors at stations 3 and 4, particularly at night. Interestingly, a unique peak at $24$Hz is also observed in Figure \ref{fig:PSD_all_stations}(a) and \ref{fig:PSD_all_stations}(b), which might represent the frequencies of the reflected waves.  

\subsubsection{Eigenmodes}
\begin{figure}[htb]
    \centering
    \begin{subfigure}[b]{0.5\textwidth}
       \adjustbox{valign=t}{\textbf{(a)}} 
        \adjustbox{valign=t}{\includegraphics[width=\textwidth]{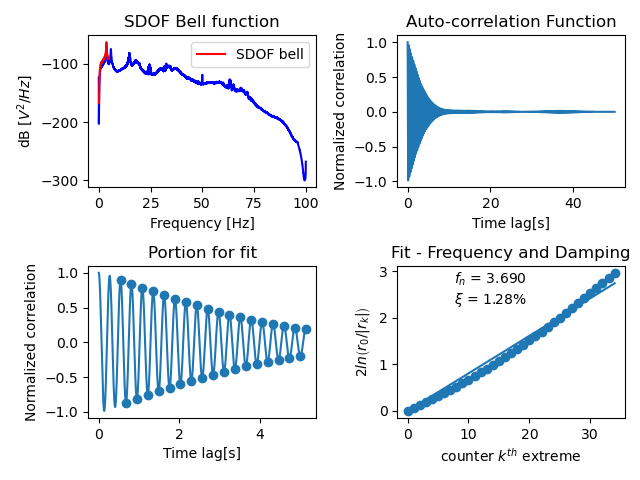}}
    \end{subfigure}
    \begin{subfigure}[b]{0.5\textwidth}
       \adjustbox{valign=t}{\textbf{(b)}} 
        \adjustbox{valign=t}{\includegraphics[width=\textwidth]{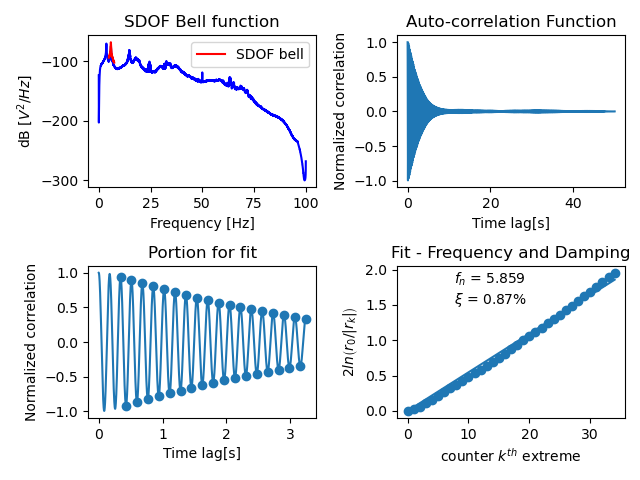}}
    \end{subfigure}
	\caption{The identification of eigenfrequencies, damping ratio and peak normalized displacement using FSDD method for (a) first mode and (ii) second mode of the BLEIB structure. Same approach is followed for the subsequenct modes.}
	\label{fig:PyOMA}
\end{figure}

\begin{figure*}[htb]
    \centering
    \includegraphics[width = 0.8\textwidth]{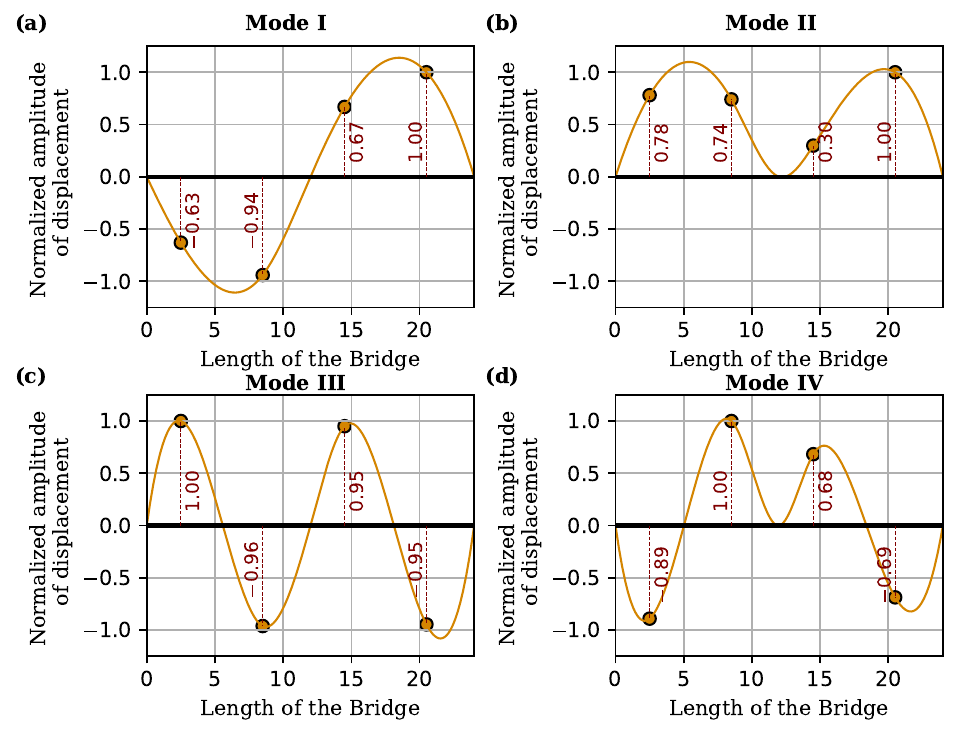}
    \caption{Eigenmodes obtained for the \textbf{(a)}: first mode, \textbf{(b)}: second mode, \textbf{(c)}: third mode,  and \textbf{(d)}: fourth mode using the ambient displacements recorded in the Z-direction.}
    \label{fig:mode_shape_TC_passive}
\end{figure*}

Once the eigenfrequencies of the bridge are obtained, the next step is to determine the Eigenmodes of the structure using the ambient translational and rotational motions. For this purpose, the Python tool PyOMA, developed by \cite{Pasca2022} is used where the FSDD numerical technique has been implemented in the present study. FSDD method is a third generation FDD (Frequency domain decomposition) method which provides accurate estimates of the modal frequencies and damping ratios for closely spaced modes \cite{Zhang2010}. As the BLEIB structure is instrumented at four locations, the recorded data at each of these is used to determine the relative displacement of the four points w.r.t each other. The PyOMA method identifies the first eigenfrequency as $3.69$Hz and the damping ratio at the first mode as $1.28\%$ (Figure\ref{fig:PyOMA}(a)). $5.859$Hz and $0.87\%$ are identified as the eigenfrequency and damping ratio respectively for the second mode (Figure\ref{fig:PyOMA}(a)). 
The third and fourth eigenfrequencies and the corresponding damping ratios are also determined using the same method. It should be noted that these eigenfrequencies are same as those observed in the PSD curves shown in Figure \ref{fig:PSD_all}(e) The numerical method only provides the relative normalized displacement of the bridge at the four locations where seismic instruments are installed. So, knowing the boundary condition that displacements are zero at the three supports, the seven points are joined by a cubic-spline smoothened curve shown in Figure \ref{fig:mode_shape_TC_passive}. Figure \ref{fig:mode_shape_TC_passive}(a), \ref{fig:mode_shape_TC_passive}(b), \ref{fig:mode_shape_TC_passive}(c) and \ref{fig:mode_shape_TC_passive}(d) describe the Mode I, Mode II, Mode III, and Mode IV obtained from the ambient noise recorded by TC sensors. 
It is observed from the figure that if the BLEIB structure is excited at its first and second natural frequency, the structure is expected to show maximum displacement at the center of both spans. However, the displacements are zero at these locations when the structure is excited at its third and fourth natural frequency. Also, each span shows two locations of maximum displacement (both at one-fourth distance from the support) at these two natural frequencies. These observations from the experimental mode shape are consistent with the analytical mode shapes expected for a simply-supported bridge with two spans \cite{Yesilce2009}. 

\begin{figure}[htb]
    \centering
    \includegraphics[width = 0.5\textwidth]{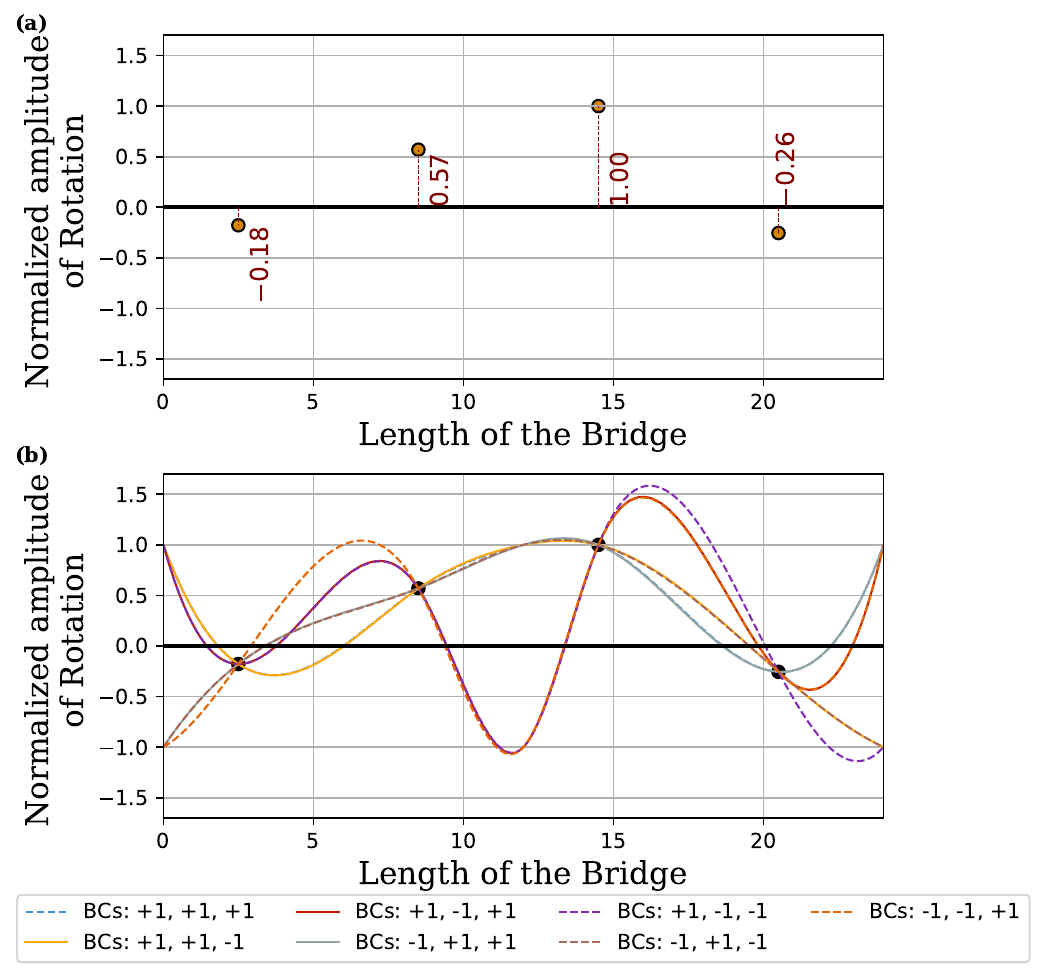}
    \caption{\textbf{(a)}: Normalized amplitude of rotations obtained at the four sensor locations when the structure is excited at its first natural frequency, \textbf{(b)}: Possible first mode shape by considering possible pairs of boundary conditions (BCs)}
    \label{fig:mode_shape_BS_passive_some}
\end{figure}

Once the translational mode shapes are determined, the next step is to determine the mode shapes from the rotational data recorded over the passive days of the experiment. Implementing the same numerical procedure as for the translational data, the relative amplitude of rotations at the four instrumented locations is obtained for the structure (Figure \ref{fig:mode_shape_BS_passive_some}(a)). As implemented for translations, the mode shapes for rotations can be obtained by connecting the four points or relative rotations and the rotations at the supports by a smooth curve. It is well known that rotations are the gradients of displacements, which are assumed zero at the supports. This would imply that the rotations are maximum at the end supports. However, it is difficult to state deterministically (i) the locations of zero rotations (ii) rotation at the intermediate support and (iii) the direction of maximum rotation at the end supports. Assuming that the rotation at the intermediate support is also positive maximum or negative maximum and not zero, there are nine pairs of boundary conditions possible for each mode shape. If the positive maximum is represented as $+1$ and negative maximum is represented as $-1$, then the boundary conditions are ($+1, +1, +1$), ($+1, +1, -1$), ($+1, -1, +1$), ($-1, +1, +1$), ($+1, -1, -1$), ($-1,+1,-1$), ($-1, -1, +1$) and ($-1, -1, -1$) . The possible rotated shape of the bridge in the first mode considering a few pairs of boundary conditions are illustrated in Figure \ref{fig:mode_shape_BS_passive_some}(b). 
It can be observed from the figure that different combinations of boundary conditions would give rise to different deformed shape of the structure resulting in contrasting regions of maximum, minimum and zero rotations. 

\begin{figure*}[htb]
    \centering
    \includegraphics[width = 0.8\textwidth]{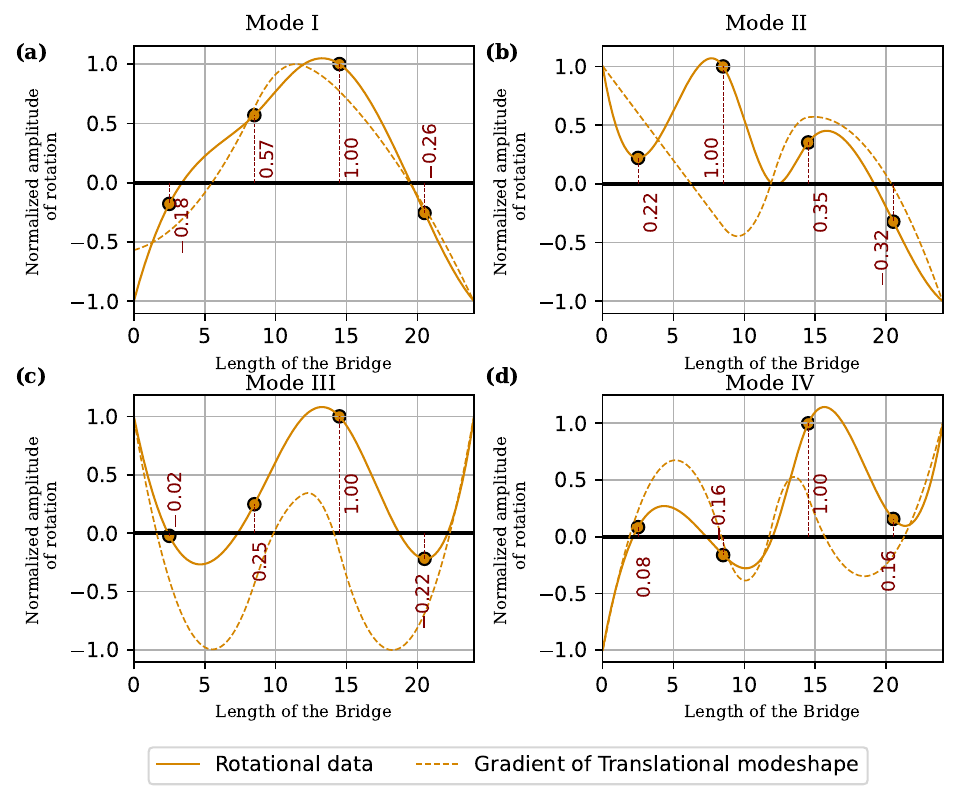}
    \caption{Eigenmodes obtained for the \textbf{(a)}: first mode, \textbf{(b)}: second mode, \textbf{(c)}: third mode, and \textbf{(d)}: fourth mode using the ambient rotations recorded about the X-direction.}
    \label{fig:mode_shape_BS_passive}
\end{figure*}

One way to determine the sign of rotations at the supports is from the translational mode shapes. The gradient of translational mode shapes can provide an estimate of the rotations at the supports, i.e., the supports at which rotation will be maximum positive ($+1$) and the ones at which the rotation will be maximum negative ($-1$). Therefore, the gradient of the translational mode shape is calculated and normalized to determine the boundary conditions for the rotational mode shape (Figure \ref{fig:mode_shape_BS_passive}(a)-\ref{fig:mode_shape_BS_passive}(d)). Once the sign of the rotation at supports is known, the relative rotations are joined by a smooth curve to obtain the rotational mode shapes. Figure \ref{fig:mode_shape_BS_passive}(a), \ref{fig:mode_shape_BS_passive}(b), \ref{fig:mode_shape_BS_passive}(c) and \ref{fig:mode_shape_BS_passive}(d) show the I, II, III and IV mode shapes respectively obtained from the rotational data and are compared with the gradient of the mode shapes obtained from the translational data. 
It can be observed from the figure that except for Mode II (Figure \ref{fig:mode_shape_BS_passive}(b)), the pattern of rotational mode shapes coincides well with the gradient of translational mode shapes for Mode I, III, and IV. However, the relative rotations obtained from the modal analysis of rotational data do not match well with the gradient of the translational mode shape. This mismatch can be attributed to two possible reasons: (i) the location of rotational sensors and (ii) unknown rotational boundary conditions. As this was one of the first experiments carried out with the rotational sensors, the BS rotational sensors were collocated with the TC translational sensors to measure the rotational motions in conjunction with the translational motions. Though, in the current arrangement of sensors, the translational instruments could capture the location of maximum displacement at the first four natural frequencies, the rotational sensors could not provide this information. Also, as the boundary conditions for the rotations are undetermined, it is likely that, in the current arrangement, the sensors were placed at locations where the rotations were not significant from the perspective of structural design. This observation is evident from the low relative rotational rate recorded at the sensor locations 1, 2, and 4 in the II, III, and IV mode shapes (Figure \ref{fig:mode_shape_BS_passive}). Further experiments with more rotation instruments placed along the length of the bridge can help to understand the locations of maximum and minimum rotations and, hence, locations of damage on the bridge.  

\subsection{Experimental Modal Analysis}
Experimental modal analysis (EMA) is another widely used method in civil engineering to determine the dynamic properties of the structure. This method not only helps in validating the analytical model to confirm the involved simplifications and assumptions but also helps in the identification of modal parameters. EMA data can be used in dynamic response analysis to evaluate how a structure will behave under different loading conditions, including seismic events, wind loads, or machinery-induced vibrations. This information is crucial for ensuring that structures can withstand anticipated dynamic forces without experiencing excessive deformations or failures. In this method, the structure is subjected to controlled dynamic forces, such as impacts, shakers etc, and the corresponding data is recorded and analyzed. As EMA involves controlled excitation, it is challenging to employ this method on real-life complex structures. However, in the current work, as the experiments were carried out on a prototype bridge, it was possible to carry out controlled active experiments. Therefore, in the next step, active experiments were carried out on the bridge, which included (i) change in prestress levels, (ii) applying additional loads, and (iii) hammer hits. The hydraulic jacks attached to the pretension cables are used to decrease the prestress from $450$kN to $200$kN and then increase back to $450$kN in steps of $50$kN each. At each prestress level, data was recorded for the ambient noise, change in the additional loads, and three consecutive hammer hits. This set of active experiments was carried out over a duration of 2 days, and the conventional TC and BS sensors in addition to IMU50 sensors continuously recorded the data. 

\begin{figure*}[htb!]
    \centering
    \includegraphics[width = 0.9\textwidth]{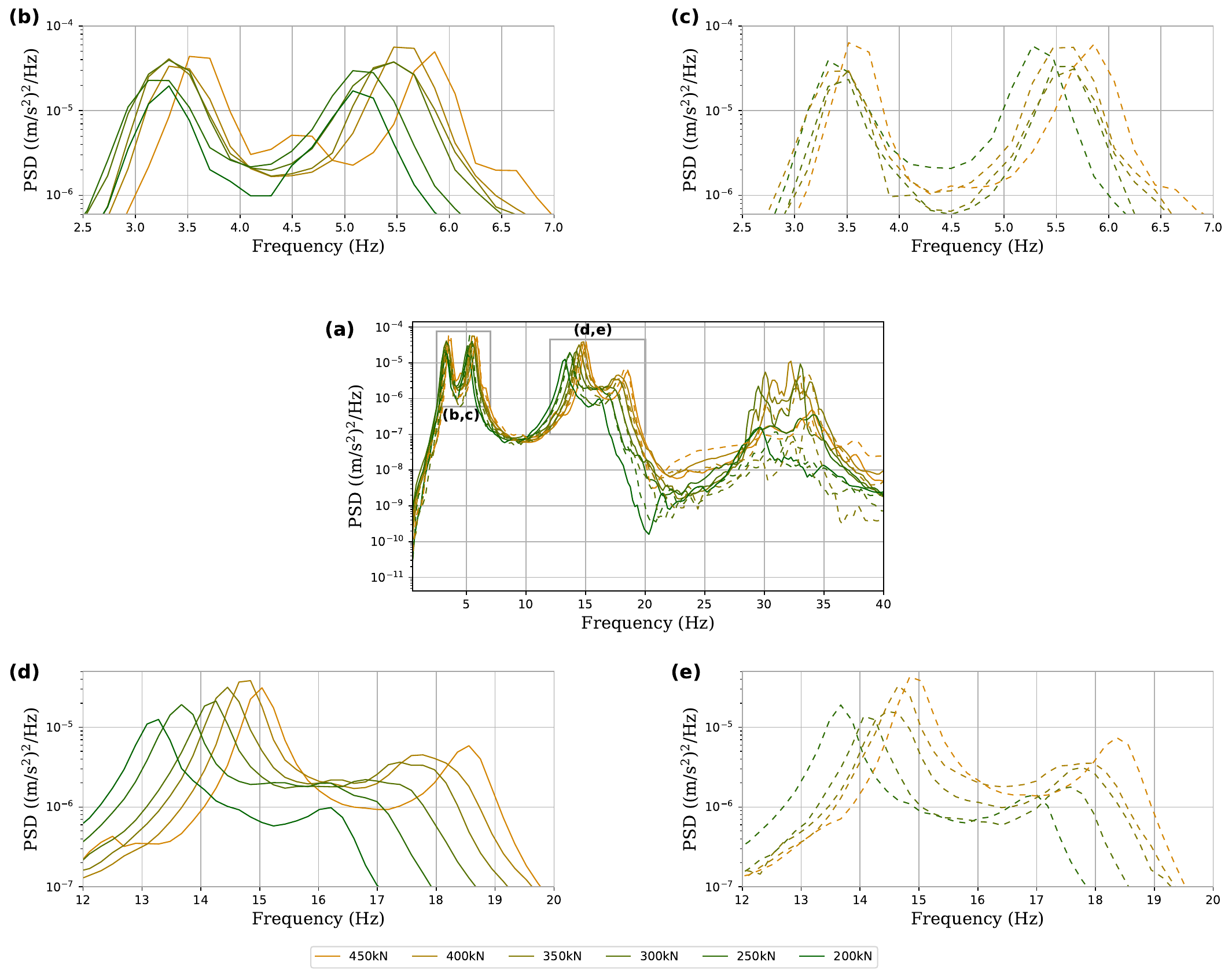}
    \caption{\textbf{(a)}: Variation of PSD function w.r.t frequency for the data recorded by TC on the active days of the bridge. \textbf{(b)}: Zoomed-in image of first and second modal frequencies obtained at different prestress levels while reducing the prestress from $450$kN to $200$kN. \textbf{(c)}: Zoomed-in image of first and second modal frequencies obtained at different prestress levels while increasing the prestress from $250$kN to $450$kN. \textbf{(d)}: Zoomed-in image of third and fourth modal frequencies obtained at different prestress levels while reducing the prestress from $450$kN to $200$kN. \textbf{(e)}: Zoomed-in image of third and fourth modal frequencies obtained at different prestress levels while increasing the prestress from $250$kN to $450$kN.}
    \label{fig:active_detailed}
\end{figure*}

\begin{figure*}[htb!]
    \centering
    \includegraphics[width = 0.8\textwidth]{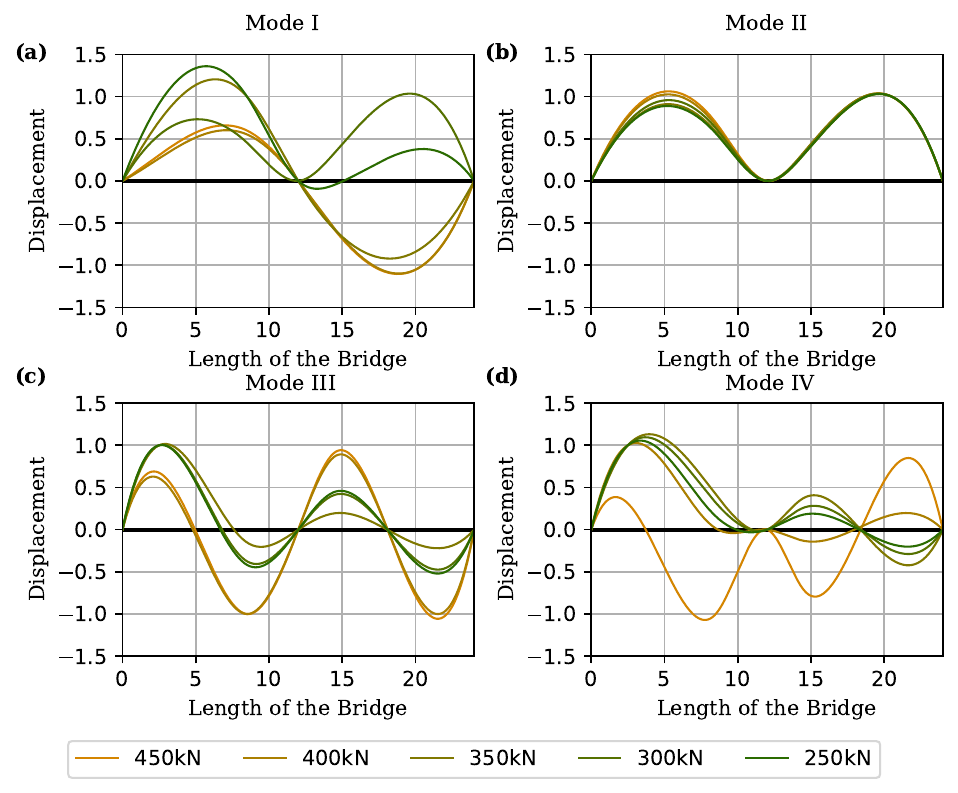}
    \caption{Eigenmodes obtained for the \textbf{(a)}: first mode, \textbf{(b)}: second mode, \textbf{(c)}: third mode, and \textbf{(d)}: fourth mode using the data recorded by the TC sensors in the Z-direction on active days of the experiment.}
    \label{fig:mode_shape_TC_active}
\end{figure*}

In order to analyze the effect of prestress on the dynamic properties of the structure, the recorded data at each prestress level is separated, bandpass filtered, and detrended to obtain the variation of PSD w.r.t frequency. Therefore, the recorded data of each prestress level captures not only the effect of prestress but also the effect of additional load and hammer hits on the structure. Figure \ref{fig:active_detailed}(a) describes the variation of PSD wrt frequency for the data recorded at each prestress level by the TC sensor at Station 4.   
The solid lines in Figure \ref{fig:active_detailed}(a) describe the PSD of recorded data while decreasing the prestress from $450$kN to $200$kN, and the dashed lines describe the data recorded while increasing the prestress from $250$kN to $450$kN. Figure \ref{fig:active_detailed}(b) and \ref{fig:active_detailed}(c) show the first two modal frequencies obtained during decreasing and increasing prestress, respectively. Similarly, Figure \ref{fig:active_detailed}(d) and \ref{fig:active_detailed}(e) show the third and fourth modal frequencies obtained during decreasing and increasing prestress, respectively. It can be observed from Figure \ref{fig:active_detailed}(b) that the first modal frequency consistently shifts leftwards from $3.5$Hz for $450$kN prestress to $3.2$Hz for $200$kN prestress level. Similarly, the second modal frequency decreases continuously from $5.6$Hz for $450$kN prestress value to $5$Hz for $200$kN prestress level. This shift in the modal frequencies is predominantly observed at the third peak frequency (Figure \ref{fig:active_detailed}(d)) where at $450$kN prestress, the peak is observed at $15$Hz while it is $13.1$Hz for $200$kN prestress. A shift from $18.5$Hz to $16.2$Hz is also observed while decreasing the prestress from $450$kN to $200$kN at the fourth peak frequency (Figure \ref{fig:active_detailed}(d)). As the prestress is increased from $200$kN to $450$kN, the four modal frequencies show an increasing trend (Figure \ref{fig:active_detailed}(c) and \ref{fig:active_detailed}(e)). The same analysis is carried out for the TC sensors placed at locations 1, 2, and 3 to determine the natural frequencies at each prestress level. Figure \ref{fig:active_comp_TC_all_locations} shows the change in the first four natural frequencies of the BLEIB structure with the decrease and increase in prestress at the four sensor locations. It is evident from the figure that the change in natural frequencies of the bridge with the change in prestress is well captured by the TC sensors placed at locations 2, 3, and 4. However, the sensor at station 1 fails to capture the fourth natural frequency. The location of station 1 is farthest from the additional loads, and hammer hits on the bridge can be a possible reason for this behavior. Also, the decreasing and increasing trend, here onwards referred to as the `V' trend, in the natural frequencies is less prominently observed for the first two modes as compared to the third and fourth modes of the bridge. The change in prestress in the BLEIB structure changes the stiffness of the structure, which in turn gives rise to the V-trend in natural frequencies. Once the natural frequencies are determined, eigenmodes of the BLEIB structure are evaluated in the next step. Following the same procedure as described in the previous section, the first four eigenmodes are obtained at each prestress level of the bridge (Figure\ref{fig:mode_shape_TC_active}(a)-\ref{fig:mode_shape_TC_active}(d)).  
The figure shows that along with the eigenfrequencies, the eigenmodes also change with change in the prestress in the bridge. Mode II (Figure \ref{fig:mode_shape_TC_active}(b)) and Mode III (Figure\ref{fig:mode_shape_TC_active}(c)) show a consistent decrease in the normalized peak displacements obtained at different prestress levels while maintaining same deformed shape of the bridge. However, in the case of Mode I (Figure \ref{fig:mode_shape_TC_active}(a)) and Mode IV (Figure \ref{fig:mode_shape_TC_active}(d)), the deformed shape of the bridge also changes with the change in prestress levels due to possible variation in the stiffness of the structure. This analysis indicates that, for prestressed concrete structures, the dynamic analysis plays a further crucial role as the behaviour of the structure changes drastically with changes in the prestress.  

\subsection{Performance comparison of IMU50 sensors}
\begin{figure}
	\centering
    \begin{subfigure}[b]{0.5\textwidth}
         \includegraphics[width=\textwidth]{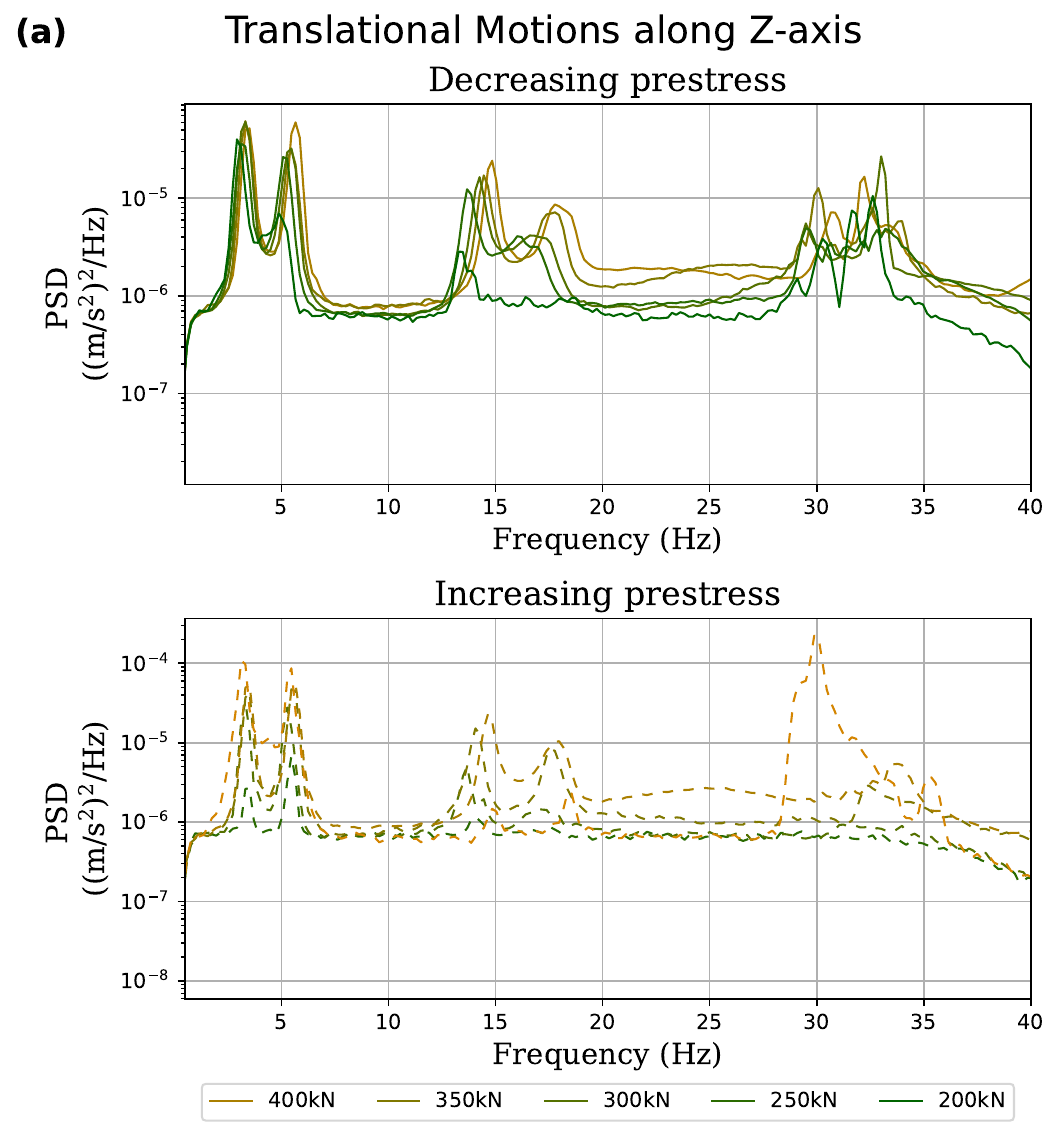}
     \end{subfigure}\\
     \begin{subfigure}[b]{0.5\textwidth}
         \includegraphics[width=\textwidth]{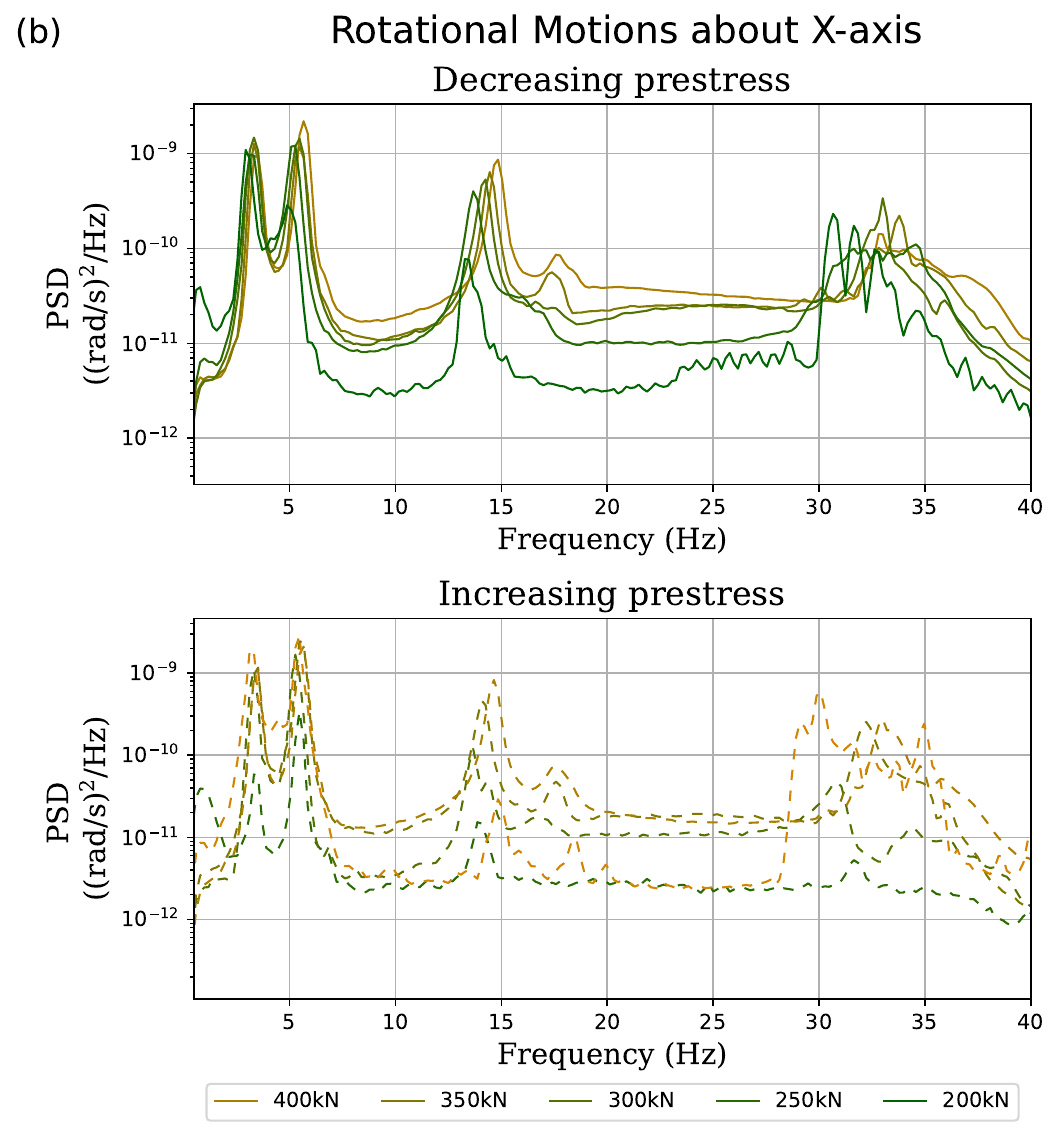}
     \end{subfigure}
    \caption{Variation of PSD w.r.t frequency observed at Station 3 using the IMU50 sensor \textbf{(a)}: translational data recorded by the IMU50 sensor along the \textit{z-} axis (top panel: decreasing prestress and bottom panel: increasing prestress) and \textbf{(b)}: rotational data recorded by the IMU50 sensor about the \textit{x-} axis (top panel: decreasing prestress and bottom panel: increasing prestress).}
    \label{fig:PSD_IMU50_detail}
\end{figure}

\begin{figure*}
    \centering
    \includegraphics[width = 0.9\textwidth]{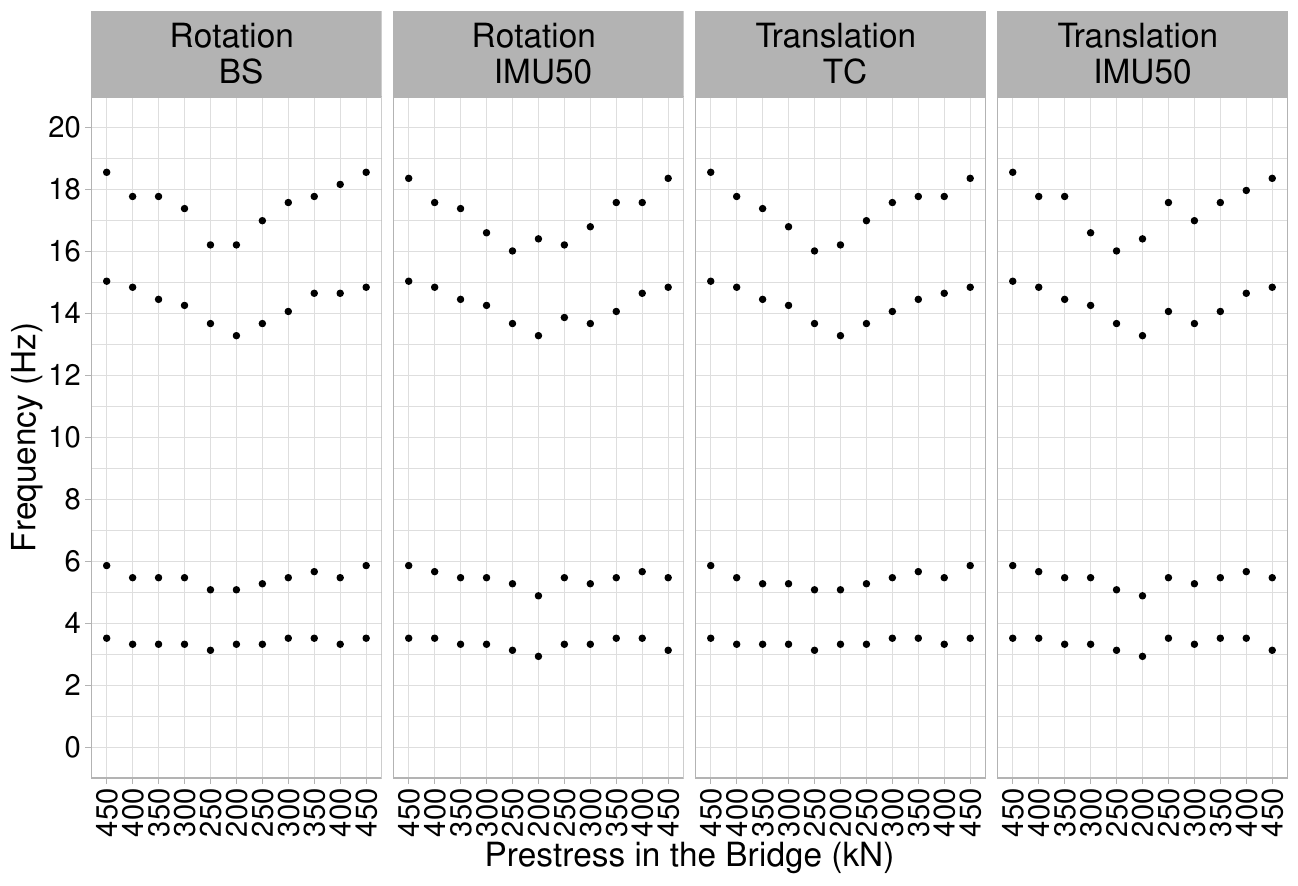}
    \caption{Variation of first four eigenfrequencies of the BLEIB structure under different prestress conditions obtained from conventional TC and BS sensor and 6C IMU50 sensors placed at location 3.}
    \label{fig:active_comp_IMU50_TC}
\end{figure*}

As one of the critical goals of the experiment was to determine the suitability of newly developed IMU50 sensors for structural health monitoring, it was necessary to first compare its performance with those of the conventional sensors. Therefore, in addition to the TC and BS sensors, 4 IMU50 sensors were also placed at the same location on the BLEIB structure for the two active days of the experiment. As done for the data obtained from conventional TC and BS sensors, the recorded data at each prestress level is separated, bandpass filtered, and detrended to obtain the variation of PSD w.r.t frequency. Figure \ref{fig:PSD_IMU50_detail}(a) and \ref{fig:PSD_IMU50_detail}(b) shows the variation of PSD of translational and rotational data respectively w.r.t frequency obtained from the IMU50 sensors at different prestress levels.   
It is observed from the figure that the magnitudes of acceleration and rotation rate recorded by IMU50 sensors compare well with those recorded by the conventional TC and BS sensors, respectively. Interestingly, it is also observed that the sensor is capable to capture the shift in the eigenfrequencies with change in prestress. The higher frequencies of the structure are also captured by the sensor, which makes it suitable for structural health monitoring purposes. In the next step, the eigenfrequencies obtained from IMU50 sensor are compared with those obtained from the conventional TC and BS sensors.  Figure \ref{fig:active_comp_IMU50_TC} shows the comparison of the first four natural frequencies obtained from the translational and rotational data recorded by the IMU50 sensor at location 3 with the conventional TC and BS sensors.
It is evident from the figure that the V-trend of natural frequencies obtained from the rotations recorded by the IMU50 sensors compares very well with that obtained from the conventional BS sensor. While comparing the V-trends of translational data, there is a slight off-peak observed at the modal frequencies at the $250$kN prestress level. Nonetheless, the IMU50 sensors show a fair match with the modal frequencies obtained from the conventional TC sensors for different prestress levels. This observations paves way to use IMU50 sensors further in the domain of structural health monitoring. As the sensors are compact, easy to mount, reliable and measure six-components directly, there use in the monitoring of civil engineering structures will certainly provide better understanding of rotational motions on the building design.   

\section{Conclusions}\label{sec5}
Civil engineers worldwide strive to design earthquake-resistant infrastructure to minimize the damage due to earthquakes. Using advanced instruments in the recent past, seismologists have determined that seismic waves cause both translations and rotations. This development has put forth a new challenge for structural engineers to determine if the existing design strategies can withstand damage due to rotational ground motions. In this regard, the present paper makes a novel attempt to use 6-DOF data to understand the dynamic properties of a structure. A $24$m long reinforced concrete prestressed bridge structure is instrumented using (i) conventional translation and rotation sensors and (ii) newly developed 6-DOF IMU50 sensors for a total duration of 18 days. During this experiment, the conventional TC and BS sensors are used to record ambient data under passive operating conditions for 16 days. The bridge is subjected to external loads and changes in prestress for two days, recorded by the IMU50 and conventional sensors. It is observed that the IMU50 sensors show promising results compared to conventional sensors and can prove helpful for structural health monitoring purposes. In order to understand the location of maximum translations and rotations along the length of the bridge, the eigenfrequencies and eigenmodes of the bridge structure are determined by employing OMA and EMA techniques. It is concluded that, under the ambient noise conditions, the translational motions along the vertical direction and the rotational motions about the transverse direction of the bridge capture the maximum number of natural frequencies. Also, the number of natural frequencies captured is independent of the number and location of these instruments on the bridge. Under experimental conditions, the eigenfrequencies shift to a lower/higher value depending on whether the prestress is decreased/increased. This shift in the eigenfrequencies is observed at all four modes and in both the translations and rotations. This behavior of the bridge under experimental conditions indicates the possibility of a shift in the region of maximum translation/rotation with a change in the prestress of the bridge. The eigenmodes of the bridge are analyzed using both translation and rotation records. The mode shapes obtained from the translation data are in accordance with the analytical mode shapes expected for a 2-spanned simply-supported bridge. However, it is difficult to determine the rotational mode shapes using data from just four sensors. Ambiguity in the boundary conditions of rotations at the support and locations of maximum or minimum rotations makes it challenging to determine the rotational mode shapes. However, with the growing evidence of rotations in the earthquake ground motions, there is a need to understand the behavior of rotations in structures and identify if the damage pattern of the buildings changes in the presence of rotations. Therefore, further experiments with more rotational sensors along the length of the bridge need to be carried out to help understand the effect of rotations on the overall mode shapes of the structure. 

\bibliographystyle{SageH}
\bibliography{literature_EESD_final.bib}

\begin{acks}
	The authors would like to extend their gratitude to Dr. Falk Hille and his team from BAM division 7.2, Joachim B\"{u}low and Regina Maass from UHH and Saskia from LMU for their support during the experiment.
\end{acks}

\begin{dci}
	The authors declare no potential conflict of interests.
\end{dci}

\begin{funding}
	This study is a part of the GIOTTO project funded by Bundesministerium f\"{u}r Bildung und Forschung (BMBF) in the frame of ``Fr\"{u}herkennung von Erdbeben und ihren Folgen" program (Grant No: 03G0885D). 
\end{funding}

\renewcommand\thefigure{\thesection.\arabic{figure}} 
\begin{sm}    
	\begin{figure*}[htb!]
		\centerline{\includegraphics[width=0.8\textwidth]{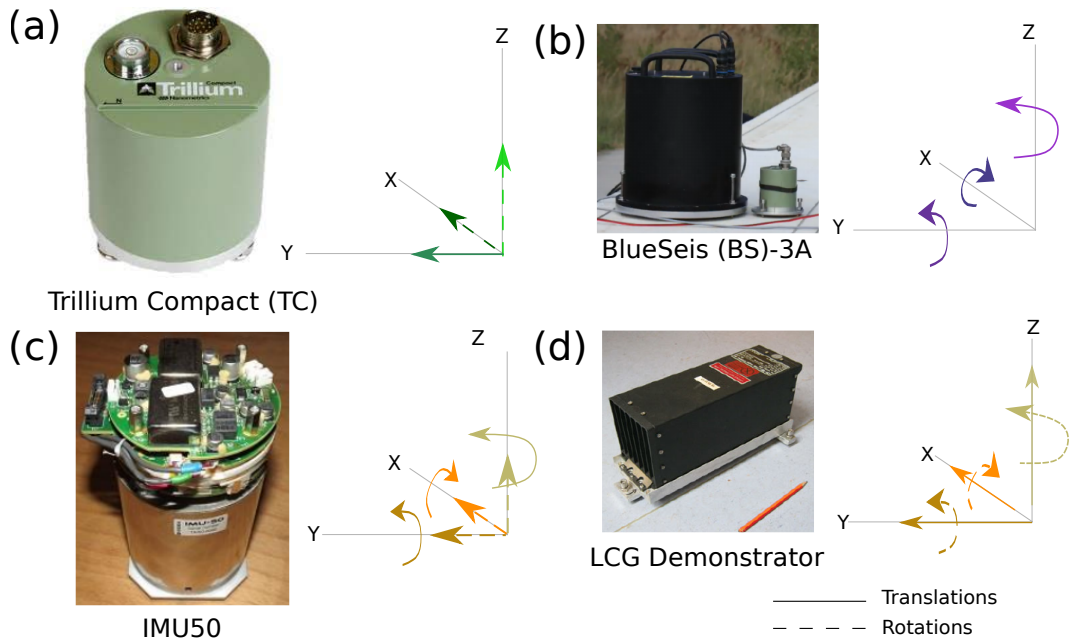}}
		\caption{Sensors used during the experiment to instrument the bridge structure (a) LCG demonstrator (b)BlueSeis-3A, (c) Trillium Compact and (d) IMU50 figure.\label{fig:instruments}}
	\end{figure*}
	
	\begin{figure*}[htb!]
		\centering
		\includegraphics[width = 0.9\textwidth]{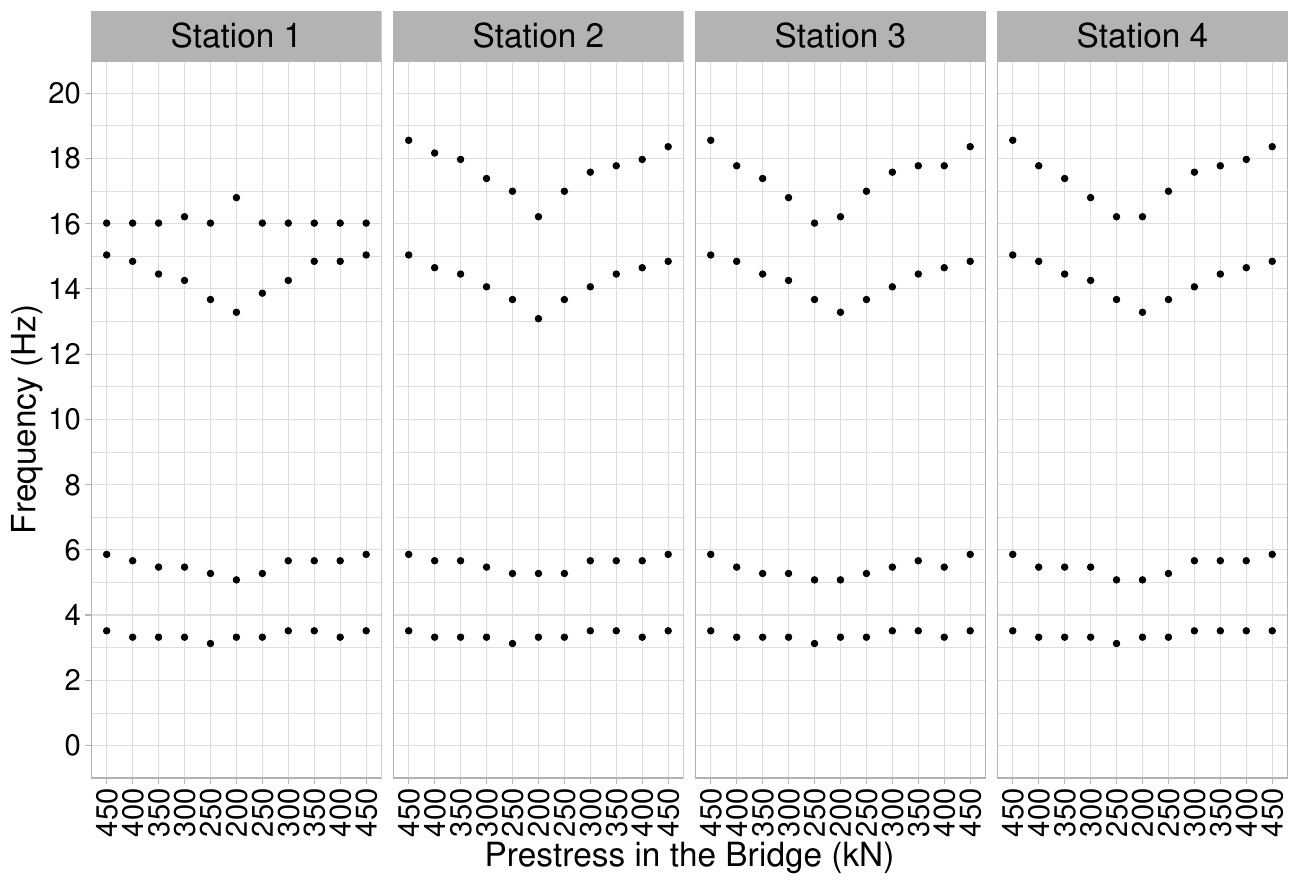}
		\caption{Variation of the natural frequencies of the BLEIB structure at the four TC sensor locations with change in the prestress levels.}
		\label{fig:active_comp_TC_all_locations}
	\end{figure*}
\end{sm}
\end{document}